\documentclass{iaus}
\usepackage{graphicx}
\usepackage{natbib}
\def \kms {km s$^{-1}$ }
\def \lsol {L$_{\odot}$ }
\def \msol {M$_{\odot}$ }
\def \kmss {km s$^{-1}$}

\def \msols {M$_{\odot}$}

\title[Understanding Arp\,220] 
{Arp\,220 - IC\,4553/4: understanding the system and diagnosing
the ISM}

\author[Baan]   
{Willem A. Baan }

\affiliation{ASTRON, Oude Hoogeveensedijk 4, 7991PD Dwingeloo, The
Netherlands \break email: baan@astron.nl }

\pubyear{2007}
\volume{242}  
\pagerange{0}
\date{16 April 2007 and in revised form ??}
 \jname{Astrophysical Masers and their
Environments} \editors{J.M. Chapman \& W.A. Baan, eds.}
\begin{document}

\maketitle

\begin{abstract}
Arp\,220 is a nearby system in final stages of galaxy merger with
powerful ongoing star-formation at and surrounding the two nuclei.
Arp\,220 was detected in HI absorption and OH Megamaser emission
and later recognized as the nearest ultra-luminous infrared galaxy
also showing powerful molecular and X-ray emissions. In this paper
we review the available radio and mm-wave observational data of
Arp\,220 in order to obtain an integrated picture of the dense
interstellar medium that forms the location of the powerful
star-formation at the two nuclei.

\end{abstract}

\firstsection 

\section{OH MM and ULIRGs}

The galaxy IC\,4335, later referred to as Arp\,220, was recognized
early as a merging and strongly interacting system with a blue
nucleus by \cite{Arp66} and \cite{Nilson73}. The detection of
strong HI absorption against the radio continuum of the nuclear
starburst \citep{Mirabel82} led to a search for corresponding OH
absorption, which resulted in the detection of powerful OH maser
emission later classified as OH MegaMaser emission (OH MM)
\citep{BaanWH82}. The extraordinary properties of Arp\,220 were
confirmed by the IRAS far-infrared (FIR) data \citep{SandersEA88},
which made Arp\,220 the prototype Ultra-Luminous FIR Galaxy
(ULIRG). The enhanced emission line strength of the molecular
constituents were recognized and the CO lines were used to map the
nuclear regions of ULIRGs \citep{SolomonEA97, DownesS98,
SakamotoEA99}. More recently high-density molecular tracers have
been studied in order to diagnose the molecular medium
\citep{AaltoSWH07, BaanEA07, Gracia-CarpioEA06, GreveEA07}.

Arp\,220 experiences a powerful burst of starformation at each of
the nuclei triggered by the ongoing merger, which results in the
FIR prominence and multiple radio SNRs at each of the nuclei
\citep{LonsdaleEA06} and SB-related hard X-ray emission
\citep{IwasawaEA05}. Arp\,220 is a recent example of a short-lived
burst of assembly and nuclear activity that are common at
redshifts of 2.5 and that defines the characteristics of massive
galaxies in the nearby universe. Arp\,220 is a template for high
redshift active galaxies.

\section{Dynamics of Arp\,220 and where is what?}

The two nuclei of Arp\,220 are separated by only 0.97" (365 pc) as
determined from their radio positions. They appear not yet
severely deformed at this advanced stage of interaction and have a
relatively small velocity difference. Therefore, the two mass
centers are still sufficiently far away from each other and need
to be one behind the other. Only the high-density tracer emission
lines accurately trace the nuclear ISM of the two nuclei having
systemic velocities of 5365 \kms for the western nucleus and 5682
\kms for the eastern nucleus (see section \ref{sec:HDgas}).

\begin{figure}
\includegraphics[height=2.5in]{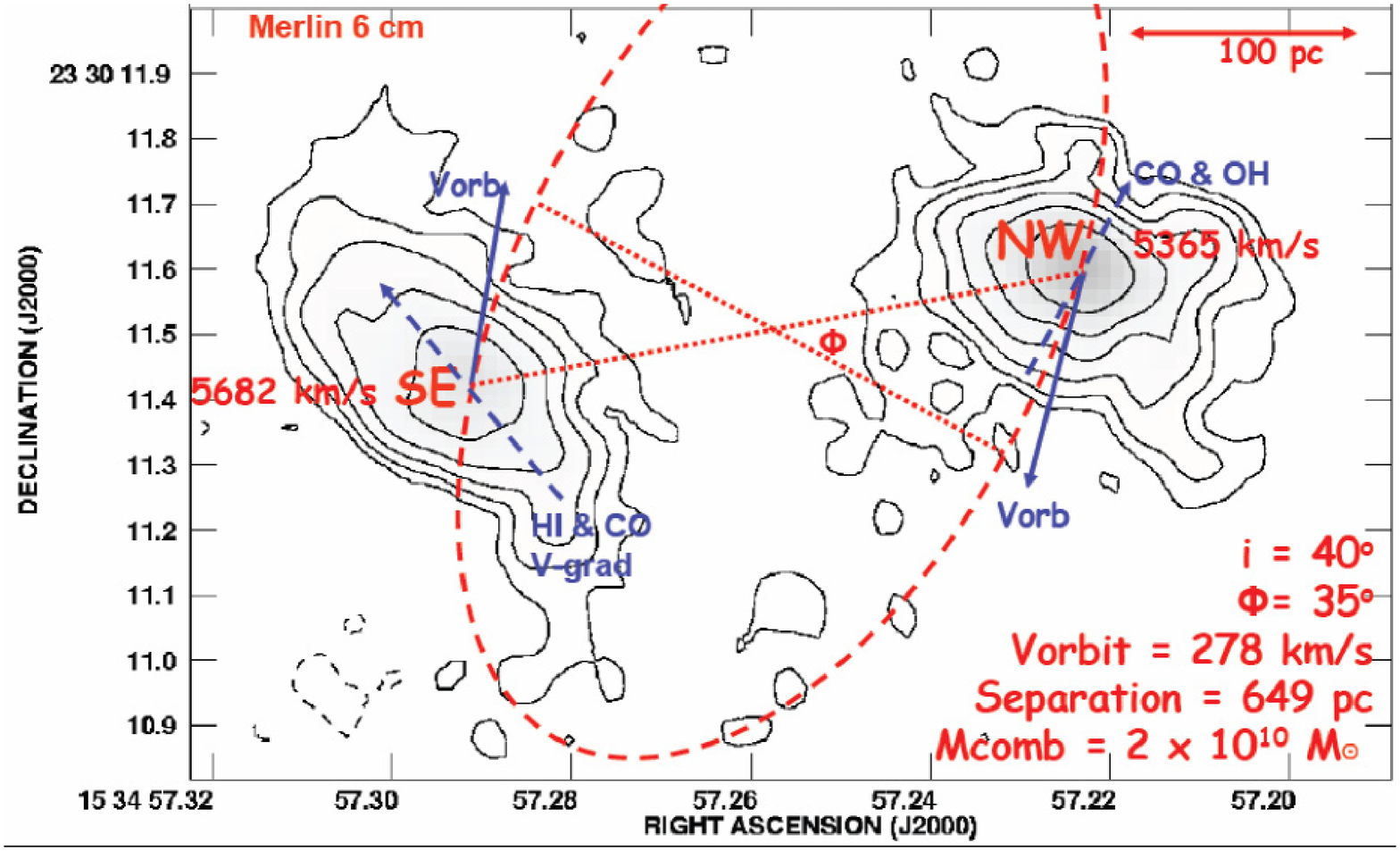}
\caption{Orbital dynamics of Arp 220. The north-western nucleus is
located in front of the south-eastern nucleus. The systemic
velocities of the two equal-mass nuclei are indicated and their
direction of motion in an orbital plane inclined by 40$^o$. The
orbital velocity ($V_{orb}$) is 278 \kmss. The combined mass of
the nuclei is based on an observed stellar and gas mass. The
directions of the observed velocity gradients of the CO and OH
molecular gas in the nuclei are indicated. The MERLIN 6cm
continuum map is from \cite{BaanCK07}. }\label{fig:dynamics}
\end{figure}

The systemic velocities of the nuclear regions obtained from the
high-density tracer emission lines may be used for dynamic
modelling of the system. Similar models were presented earlier
using H$_2$CO\,(1-0) and CO\,(2-1) molecular data \citep{BaanH95,
DownesS98, ScovilleYB97}. Reconsideration of the available
continuum, CO, HI absorption, and OH and H$_2$CO emission data
gives slightly different picture of the dynamics of the system
\citep{BaanCK07}. The eastern nucleus lies behind the western
nucleus, which would result in the observed larger column
densities in front of the eastern nucleus \citep{IwasawaEA05} and
a relatively lower velocity for the western nucleus. Using the
CO\,(2-1) signature, the orbital plane is inclined by 40$^o$
\citep{ScovilleYB97}. The combined mass of lower-density molecular
gas in CO, the high-density gas in tracers such as HCN, and the
stellar component is of the order of 0.6 - 2.0 x 10$^{10}$ \msol
\citep{SakamotoEA99,GreveEA07}. The dynamics of the system may
follow  from two equal-mass nuclei with a spatial separation of
650 pc (Fig. \ref{fig:dynamics}).

Near-infrared imaging of Arp\,220 using the NICMOS instrument (1.6
$\mu$m) onboard Hubble ST confirm the presence of high column
densities in the system \citep{ScovilleEA00}. The visual
attenuation is 17 and 24 magnitudes for the East and West nuclei
assuming a young stellar population and 13 and 20 magnitudes for
an old population (note: one magnitude corresponds to 1.8 x
10$^{21}$ cm$^{-2}$ H). The NIR emission actually peaks north of
each of the nuclei; there are areas where virtually no radiation
can escape, even at 2.2 $\mu$m. Reversing the symmetry of the
model could place the eastern nucleus in front of the western
nucleus, but would not explain the presence of OH emitting gas at
the velocity of the western nucleus in front of the eastern
nucleus (see section \ref{sec:OHem}).

The hard X-ray (3 - 7 KeV) signature of Arp\,220 as observed by
Chandra shows two separate (extended) nuclei and there is an
extended nebula at 2 keV \citep{IwasawaEA05}. The hard X-ray
continuum is due to the integrated X-ray emission from binary
systems and is a factor 10 too low for the observed SFR. There are
plasma features associated with the ionized gas and a Fe K$\alpha$
feature at 6.5 keV. The obscuration of the X-rays required $N_H$ =
10$^{22-23}$ cm$^{-2}$ (Note: $N_H$ = 10$^{22}$ cm$^{-2}$ is
enough for keV absorption).

\section{Modelling the starburst}
The power generation for the FIR luminosity of 10$^{12.2}$ \lsol
is dominated by starburst activity for more than 60\% in the radio
and more than 90\% in the near-infrared \citep{BaanK06,
GenzelEA98}. The Arp\,220 starburst is a scaled-up version of the
M\,82 activity where also Type II supernova and remnants of
massive progenitor stars interact with the ISM. Modelling of the
embedded starbursts suggests a star-formation rate SFR = 26
L$_{IR,11}$ \msol yr$^{-1}$ = 340 \msol yr$^{-1}$ and a supernova
rate SNR = 0.2 L$_{IR,11}$ yr$^{-1}$ = 2.8 yr$^{-1}$ for Arp 220
\citep{ElsonFF89}. \cite{LonsdaleEA06} have observed a SNR of
4$\pm$2 yr$^{-1}$ and find the rate in the western nucleus to be
three times higher than in the eastern nucleus. The mass injection
rate for the nuclei of Arp\,220 is estimated at 53 \msol
yr$^{-1}$. Recent evaluation of the radio data suggests a
top-heavy initial mass function (IMF) or a short starburst with a
duration of 3 x 10$^5$ yr with a SFR $\approx$ 10$^3$ \msol
yr$^{-1}$ \citep{ParraEA07,LonsdaleEA06}.

The starburst in Arp\,220 must be close to its peak in activity.
Much of the molecular ISM appears still intact but in an excited
state as evidenced from the warm high-density molecular material
(see section \ref{sec:HDgas} below). The line ratios of prominent
high-density tracers suggest that the molecular ISM is (or is at
the verge of) being depleted and destroyed by the ongoing
star-formation process \citep[see section \ref{sec:HDgas}
and][]{LoenenBS07}. There is evidence that destructive mechanical
feedback due to supernovae and high-velocity outflows seen in OH
emission are affecting the molecular ISM (section
\ref{sec:OHoutflow}). Mechanical feedback due to a violent
starburst and radiative feedback on the molecular chemistry could
cut off the ongoing star formation, leaving a single (first)
generation of stars with a top-heavy IMF.

\section{Imaging the OH emission lines}\label{sec:OHem}
Knowledge of the systemic velocities of the high-density gas at
the two nuclei (see section \ref{sec:HDgas}) also helps to
understand the complex OH groundstate line emission signature. Two
pairs of (1665 and 1667 MHz) OH emission lines are offset by 317
\kms and have a 351 \kms separation within the pair. Therefore,
triple emission lines are seen: the 1667 MHz West line, a
convolved 1667 MHz East and 1665 MHz West line, and a 1665 MHz
East line. However, also emission in the 1667 MHz line at the
systemic velocity of West is seen at the East location \cite[see
VLA-A and VLBI data by][]{BaanH87,RovilosEA03}. The gas
responsible for this 1667 MHz emission at the eastern nucleus
cannot be foreground gas trailing from the western nucleus. This
is also not solved by placing the eastern nucleus in the
foreground in the dynamical picture of Fig. \ref{fig:dynamics}.
The nature of all emissions and further verification of the
dynamical scenario for the two nuclei may be deduced from a study
of the lower-gain diffuse emission lines.

In addition to the three emission components due to two line
pairs, there is a blue-shifted component that is evidence for
outflow in the nuclear region \citep{BaanHH89}. The blue-shifted
wing of the 1667 MHz line in Arp\,220 extends by 1145 \kmss, while
the 1720 MHz line extends by 950 \kmss, although this second
outflow wing may be less-pronounced (Ghosh \& Salter 2007,
personal communication).

\begin{figure}
\includegraphics[height=5in, angle=-90]{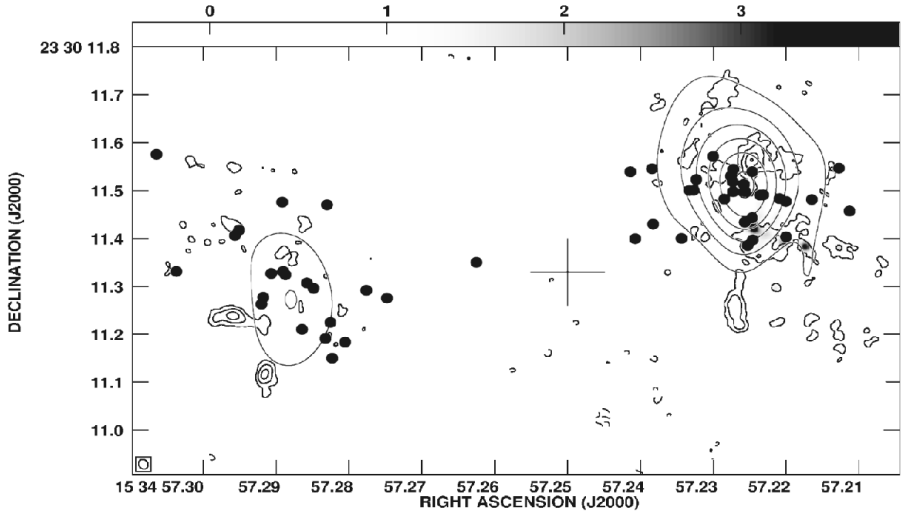}
\caption{The OH emission at the two nuclei of Arp\,220. The
high-resolution EVN map of the 1667 MHz emission from
\cite{RovilosEA03} has been overlayed with the map of supernovae
in the two nuclei from \cite{ParraEA07}. Continuum contours
indicate the location of the radio nuclei. The superposition is a
best effort. (0.1" = 35 pc)} \label{fig:OHemission}
\end{figure}

The OH emission in Arp\,220 at both nuclei encompasses the
high-resolution radio continuum structure and the high-density ISM
where the star-formation is taking place. At MERLIN resolution
(0.17" = 61 pc) the 1667 MHz OH emission at the eastern nucleus is
centrally peaked and has an extent of 170 pc along the velocity
gradient \citep{RovilosEA03}, which agrees with the extent of the
high-brightness components found at EVN resolution (0.016") (see
Fig. \ref{fig:OHemission}). Similarly the emission at the Western
nucleus shows a extended (1.2" = 420 pc) double (torus-like)
structure straddling the continuum nucleus \citep[Fig.
\ref{fig:OHemission};][]{RovilosEA03} with clumpy high-brightness
components only in the central 0.6" region. At both nuclei the
elongated/edge-on distribution of high-brightness OH components
extends beyond the starburst-related cluster of supernova remnants
in the continuum \citep{RovilosEA05, LonsdaleEA06, ParraEA07}. The
OH velocity gradients at both nuclei are similar to those in the
CO\,(2-1) data (see Fig. \ref{fig:dynamics}), which would confirm
the presence of orbiting counter-rotating disks
\citep{SakamotoEA99}.

The OH characteristics suggest that the emissions at both nuclei
originate both in the nuclear ISM and in an edge-on torus/disk
structure that surrounds the nuclear ISM with the ongoing star
formation. At the Eastern component the amplifying medium of the
major emission component is the (foreground) frontal section of
the torus at the systemic velocity. The double structure of the
Western component defines the two edge-sections of a similar thick
torus with radius of about 60 pc. The high-brightness components
emission components primarily trace structures in the nuclear
region and result from spatial superpositions of clouds and
(edge-on) shocked regions associated with the star formation. OH
emission is also seen in front of individual SNR remnants (see
Lonsdale \etal, these proceedings). The edge-on torus may serve
for further amplification. Successful modelling of observed OH
emission structures has been done for Mrk\,273 and IIIZw\,35
\citep[Parra \etal, these proceedings;][]{Kloeckner04,ParraEA05}.

\section{Clumpy low-gain amplification for OH}
The standard model for masers of low-gain amplification of radio
continuum by foreground molecular material has been suggested in
order to explain the extremely powerful emissions in OH MM
\citep{Baan85, Baan89, HenkelW90}. The spatial distribution of the
background radio continuum and the availability of a pumping agent
will allow a foreground screen to selectively amplify the
distributed continuum. Assuming a uniform screen gives a first
order estimate of the global amplification. A more realistic
non-uniform and clumpy nuclear ISM will give non-uniform
amplification in high-resolution data. Emission is found in all OH
ground state transitions in Arp\,220, while the known 5 GHz and 6
GHz lines are all in absorption in both OH MMs and absorbers
\citep{HenkelGB87, ImpellizzeriEA05}. An amplification scenario is
relevant for extragalactic and also galactic emissions of OH,
H$_2$CO, H$_2$O, and CH$_3$OH.

Evaluation of the high-resolution data suggests a gain factor of 4
to 5 relative to the background ($\tau$ = 1.4 - 1.6) for the
extended components that account for the one third of the 1667 MHz
emission \citep{RovilosEA03}. The compact components have a gain
factor ranging from 46 - 65 at the southern part of the western
nucleus to 125 at the northern part ($\tau$ = 3.8 - 4.8). Clump
superpositions or edge-on (shocked) shells need to increase the
optical depth by a factor up to three.

Pumping scenarios for the OH molecules \citep[][and Elitzur, these
proceedings]{HenkelGB87, RandellEA95, LockettE07, ParraEA05}
provide specific requirements for the physics of the ISM of the
galactic nuclei. The efficiency of the radiative pumping using the
53 and 35 µ$\mu$m FIR lines puts requirements on the FIR-SED and
requires a threshold temperature for the dust of at least 50 K,
whereas Arp\,220 has 61 K. Typically clouds of 1 pc should have an
opacity of one and a linewidth of 10 \kms for line-overlap
conditions. Pumping can be achieved across a density range of
n(H$_2$) = 10$^3$ to few 10$^4$ cm$^{-3}$. For $T_{ex}(OH)$
approximately constant, the line ratios are a function of optical
depth and give an optical depth of order 2.0, which is consistent
with observations. Simulations of the cloud superposition and the
amplification in a torus-like structure has been done by
\cite{Kloeckner04} and \cite{ParraEA05}.

\section{The high-density ISM}\label{sec:HDgas}

Diagnostics of the heavily obscured nuclear ISM can also be done
by interpreting molecular tracers to determine densities,
temperatures and the molecular chemistry. Accurate estimates of
global (integrated) properties can be obtained with LVG (large
velocity gradient) radiation transfer studies using line emission
along the energy ladder of the molecule. In addition, the chemical
evolution of molecular species combined with their radiative
transfer can be obtained by modelling X-ray and Photon (UV)
dominated regions (XDR and PDR) \cite[][also Spaans \etal\ and
Loenen \etal, these proceedings]{MeijerinkSI06, MeijerinkSI07}.

\begin{figure}
\begin{minipage}[t]{0.35\textwidth}
\includegraphics[width=\textwidth]{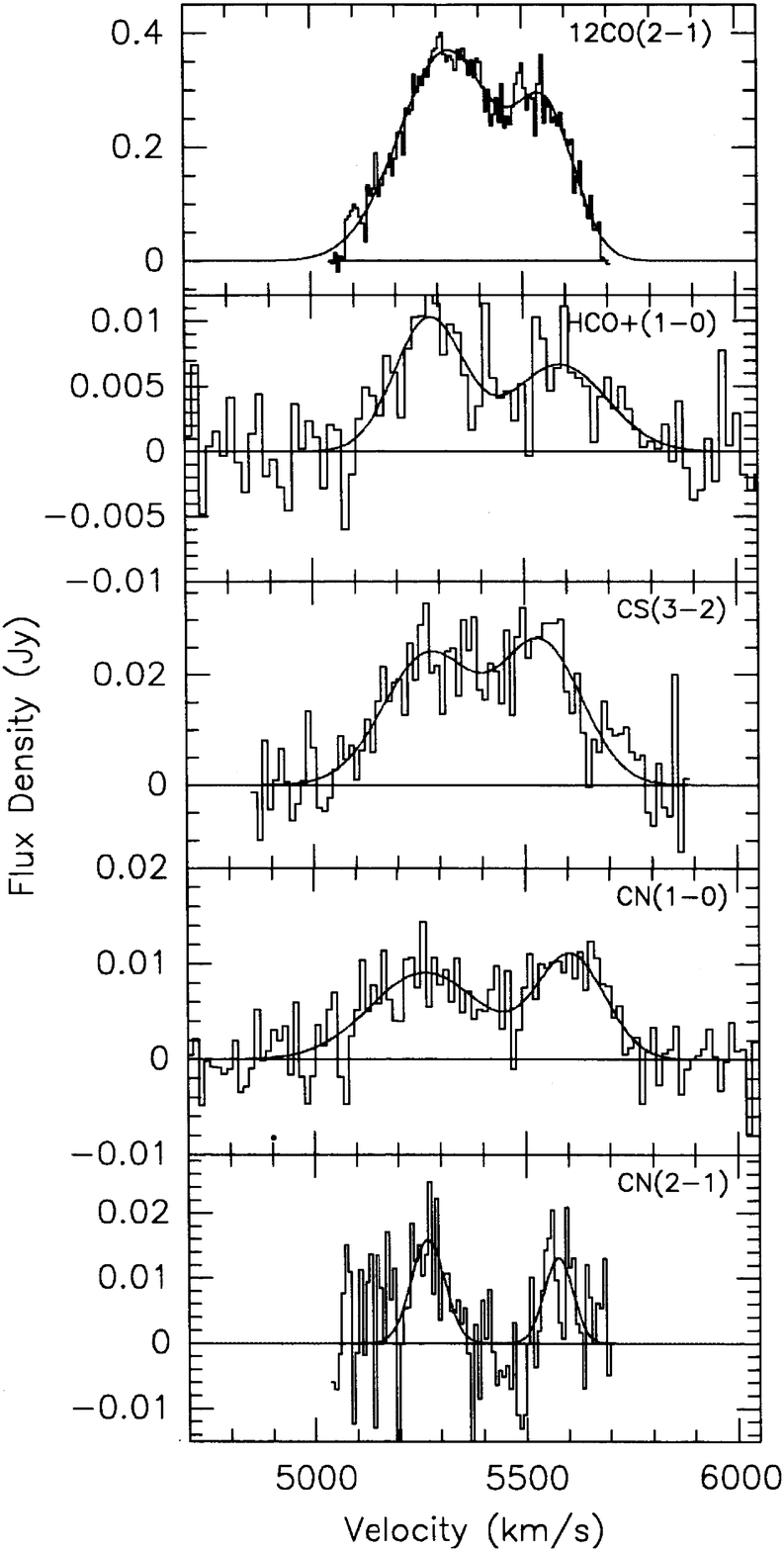}
\end{minipage}\hfill
\begin{minipage}[t]{0.60\textwidth}
\vspace{-9.2 cm}
\hspace{2cm}\includegraphics[width=0.65\textwidth]{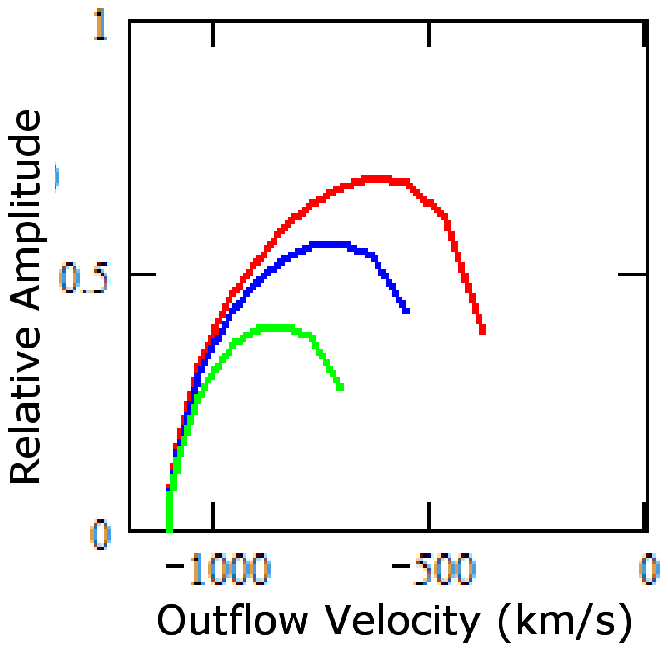}
      \caption{Spectra of emission lines of high-density tracer
      molecules in Arp\,220. From top to bottom the emission line
      spectra of CO\,(2-1), HCO$^+$\,(1-0), CS\,(3-2), CN\,(1-0),
      and CN\,(2-1). Data taken from \cite{BaanEA07}.
      \label{fig:HDlines}}
      \caption{Simulations of the amplified OH emission from
      outflows driven by supernova remnants. The three curves
      result from different covering factors for the spherical
      layers of molecular gas relative to the radio
      continuum generated by the SNR. The chosen terminal velocity
      is as observed in Arp\,220.
      \label{fig:outflowsim}}
    \end{minipage}
\end{figure}

Many molecular transitions have been detected in Arp\,220 and in
similar nearby (U)LIRGs like NGC\,6240, and Mrk\,231. Single-dish
spectra obtained with the IRAM 30\,m and SEST 15\,m telescopes of
some of the tracer lines in Arp\,220 are presented in Figure
\ref{fig:HDlines} \citep{BaanEA07}. Two features can be seen in
Fig. \ref{fig:HDlines} with systemic velocities of 5365 and 5683
\kms for the western and eastern nuclei \citep{BaanEA07}. Higher
transitions have been detected for CO, CS, CN, HCN, HNC, and
HCO$^+$ \citep{AaltoSWH07, Gracia-CarpioEA06, GreveEA07,
WiednerEA02}. Recent maps of the prominent HCN\,(4-3) line at 270
GHz display localized emission with a spatial extent similar to
that of the cluster of discrete radio sources \citep{WiednerEA07}.

The molecular ISM in ULIRGs is expected to have extreme excitation
conditions where the physical environment and the molecular
chemistry have been affected by the radiation fields and the
feedback from the star-formation process. The environment is very
different from those in Galactic SF regions. The interpretation of
extragalactic phenomena requires an extrapolation of well-resolved
Galactic phenomena to integrated and unresolved nuclear
environments. High-density components may be used to discriminate
between PDR and XDR excitation conditions \citep{BaanEA07}. In
particular, the HNC/HCN ratio has values less than unity in PDRs
and values larger than unity in XDRs. Similarly the ratios of
HCO$^+$ with HCN and HNC discriminate between the gas density due
to different critical densities of the molecules. The intensity
ratio of a high-density tracer molecule with that of a lower
density tracer, such as CO\,(1-0) that is more extended and is
less affected by the star-formation activity, shows that the
high-density medium varies strongly during the lifetime of the
activity \citep{BaanEA07}. The extreme molecular ratios for
Arp\,220 and those of other OH MM suggest extreme PDR-like
conditions for the nuclear ISM \cite[Loenen \etal\ these
proceedings and][]{LoenenBS07}.

LVG simulations suggest a two-phase ISM for the combined nuclei of
Arp\,220 with a diffuse component with n(H$_2$) = 2 x 10$^3$
cm$^{-3}$ and T$_k$ = 40 - 60 K and dense components with n(H$_2$)
= 10$^{4 - 6}$ cm$^{-3}$ and 50 - 70 K \citep{GreveEA07}. The
molecular gas is contained within self-gravitating and virialized
clouds with mass 0.6 - 2.0 x 10$^{10}$ \msols. The considerably
higher pressure in the warm high-density medium as compared with
the low-density medium will gradually cause disruption of the
medium.

\section{Signature of nuclear outflows}\label{sec:OHoutflow}

Arp\,220 is one of the OH MM galaxies showing blueshifted outflow
components for their 1667 MHz emission lines \citep{BaanHH89,
Baan07}. The observed outflow velocities vary with $L_{FIR}$ and
are likely associated with the shocked shells surrounding
supernova remnants. A continuing high-intensity starburst as found
in such ULIRGs would eventually result in merging the localized
outflows into a large-scale blowout or nuclear superwind
\cite{HeckmanAM90}. The OH outflow emission results from a
superposition shocked SNR shells propagating into the dense ISM
that amplify the embedded radio continuum. A simple simulation of
a superposition of superposed spherical outflowing shells
amplifying a spherical continuum structure provides the correct
shape and signature of the outflow (see Figure
\ref{fig:outflowsim}). The observed high velocities of the
outflows suggest that J-shock chemistry is responsible for the
molecular abundances. Molecules are destroyed and reassembled in
J-shocks for V $>$50 \kms while for V $>$300 \kms this reassembly
process happens at a slower rate. Using an energy conserving
scenario for the mechanical energy \citep[see][]{ElsonFF89}, we
find that: N$_{SN}$ R$^2_{kpc}$ V$^3_{100}$ n$_{ISM}$ $\approx$ 23
L$_{IR,11}$, which balances the total energy of the outflows from
N$_{SN}$ SNRs with outflow velocity V (in units 100 \kmss) and
with radius R (in kpc) and the injected energy L$_{IR}$ (in units
of 10$^11$ \lsol). Using L$_{IR}$ = 1.6 x 10$^{12}$ \lsol for
Arp\,220, an observed number of N$_{SN}$ $\approx$ 100, an
observed maximum velocity V(max) = 1150 \kmss, and a typical SNR
radius = 1 pc, we find that the mean ambient density of he
pre-shock material n$_{ISM}$ = 1.5 x 10$^3$ cm$^{-3}$, which is in
an acceptable range. The interpretation of these outflows clearly
requires more simulations as it provides further evidence of the
eruptive nature of the star-forming environment.

\section{Formaldehyde emission }

Formaldehyde emission has been verified in Arp\,220 as well as two
other OH MM, and an H$_2$O MM using the Arecibo and Effelsberg
telescopes \citep{BaanHU93,ArayaBH04}. At low resolution the
formaldehyde in Arp\,220 peaks at the western nucleus and has weak
emission towards the eastern nucleus and the connecting structure
\citep{BaanH95}. Recent MERLIN data shows an elongated emission
region of 80 pc at the western nucleus (resolution 0.04'' = 15 pc)
centered at 5325 \kmss, which is 40 \kms below the systemic
velocity of the western nucleus \citep{BaanCK07}. The velocity
gradient of H$_2$CO aligns with the weak gradient of the OH
emission.

The diagnostics of this emission is still limited by the
understanding of the pumping mechanism. Because the formaldehyde
emission at lower resolution mimics the 2.2 $\mu$m emission, the
IR radiation field could play a role in the pumping. A first
pumping model proposed for the excitation of the extragalactic
formaldehyde was based on pumping by the non-thermal radio
continuum, which gives small (few percent) optical depths for
densities of 10$^{2-5}$ cm$^{-3}$ \citep{BaanGH86}. While this
model could operate in extragalactic environments, it poses a
problem for small scale Galactic environments. Evaluation of
collisional and other pumping schemes for Galactic and
extragalactic environments shows that optical depths at the
relevant density range are still relatively small and may not yet
satisfy maser requirements under Galactic conditions
\citep[see][and Darling, these proceedings]{ArayaBH04, BaanHU93,
ArayaEA07}. The high- and low-resolution formaldehyde data
suggests that emission may be tracing high- and low-density gas
components in a larger (foreground) gas structure that is not
confined to the site of the star-formation itself.

\section{Conclusions}\label{sec:concl}

Radio observational tools contribute significantly to the
understanding of the nuclear medium of Arp\,220. Arp\,220 is
dynamically advanced such that the two nuclei have developed a
common gas/stellar structure in advance of the merger. The current
dynamical picture can be determined on the basis of the observed
velocities of the two nuclear ISM components. Placing the eastern
nucleus, with the highest obscuration, behind the western nucleus
on a (first approximation) circular orbit explains the large-scale
characteristics of the system but it does not yet explain OH
emission at discrepant velocities and the extended continuum
structure between the nuclei \citep[see][]{RovilosEA03}.

The observed molecular components in Arp\,220 sample a density
range from 10$^2$ to 10$^6$ cm$^{-3}$. The low-density medium with
10$^{2-3}$ cm$^{-3}$ is traced by the extended diffuse and clumpy
CO components. The CO\,(2-1) transition traces the upper end of
this range in nuclear regions enveloping the highest density
medium. The high-density constituents have densities in the range
10$^4$ to 10$^6$ cm$^{-3}$ depending on the critical density of
the molecules. LVG simulations suggest that the temperature of the
low-density tracer emissions ranges from 40-60 K, while those of
the higher density components suggest 50-70 K. This overpressure
of the high-density components results from radiative and
mechanical feedback of the star-formation process. The integrated
dust temperature from the FIR SED is about $T_d$ = 40 K, while the
local conditions in Arp\,220 with $T_d$ $>$60 K provide the
required pumping photons for the OH emission.

Both nuclear regions harbor a powerful and rapidly evolving
starburst at the point destroying its own ISM. The star-formation
rate is on the order of 340 \msol yr$^{-1}$ and occurs
predominantly in the western nucleus. The ISM in these nuclei is
still mostly intact but there is evidence of radiative and
mechanical feedback at work. The overpressure of the high-density
molecular constituents and the observed OH outflows will in time
redistribute and homogenize the nuclear ISM. Radiative feedback
also results in a PDR dominated molecular chemistry. The effects
of feedback on the molecular medium and the star-formation process
would lead to a top-heavy (truncated) IMF and lead to the early
termination of the star-formation. There is no evidence for the
presence of an AGN in the nuclear regions and any associated XDR
conditions. In addition, the observed emission due to X-ray
binaries is too weak for the level of star-formation activity. The
nuclei of Arp\,220 have become the prototype of an extreme nuclear
environment with dominant PDR conditions.

The FIR-pumped OH emission extends beyond the nuclear activity
regions and shows diffuse structures as well as clumpy and
shell-like components with densities of 10$^{2-4}$ cm$^{-3}$
\citep[see Fig. \ref{fig:OHemission}][]{RovilosEA03}. The diffuse
and clumpy H$_2$CO can be collisionally or radiatively pumped for
densities of 10$^{3-5}$ cm$^{-3}$. The SNR-driven shells seen in
the OH emission are not yet well-constrained but ambient densities
of ~10$^3$ cm$^{-3}$ are consistent. The diffuse and the
high-brightness OH components cover regions with sizes of 1 to 100
pc, and result from a distributed clumpy molecular medium
contained in two edge-on thick tori with a radius of about 60 pc
that surround the nuclear ISM at each nucleus. The highest
amplifying gains result from cloud superpositions and edge-on
shells (also seen as outflows) in the nuclear ISM. The OH and
H$_2$CO emission structures trace intermediate-density components
surrounding the highest density ISM structures co-located with the
sites of star-formation. The special conditions leading to OH and
H$_2$CO MM activity may occur during a significant part of the
duration of the starburst.

The high-resolution VLBI data on the OH and H$_2$CO emissions and
the data for high-density molecular tracers provide a
complementary picture of the nuclear ISM and the ongoing SB
activity in Arp\,220. While the emission of the high-density
tracer molecules presents the integrated properties of the nuclear
ISM and is mostly unresolved at the two nuclei, the
high-resolution data on the OH and H$_2$CO emissions provides
sufficient structural detail of the medium surrounding the sites
of star formation. The interpretation of integrated properties
depends on our ability to extrapolate from the understanding of
Galactic environments to the extreme case of unresolved nuclear
environments in starburst nuclei.


\end{document}